\documentclass[twocolumn,showpacs,preprintnumbers,
amsmath,amssymb]{revtex4}
\usepackage{graphicx}
\usepackage{verbatim}

\def\al{\alpha}
\def\be{\beta}

\def\de{\delta}

\def\eps{\varepsilon}
\def\th{\theta}
\def\la{\lambda}
\def\si{\sigma}

\def\cD{{\mathcal D}}
\def\cL{{\mathcal L}}
\def\cJ{{\mathcal J}}
\def\cM{{\mathcal M}}
\def\cN{{\mathcal N}}
\def\cO{{\mathcal O}}

\def\11{{\mathbb 1}}

\def\re{{\rm e}}
\def\ri{{\rm i}}

\def\beq{\begin{equation}}
\def\eeq{\end{equation}}
\def\bea{\begin{eqnarray}}
\def\eea{\end{eqnarray}}
\def\nn{\nonumber}

\def\rarr{\rightarrow}

\def\ri{\text{i}}

\def\dslash{\partial \hspace{-1.3ex}\slash}
\def\dslashbar{\bar\partial \hspace{-1.4ex}\slash}

\def\pslash{p \hspace{-1.0ex}\slash}

\begin{document}
\title{Neutrino loops from neutrino mixing}
\author{Adam Latosi{\'n}ski$^1$, Krzysztof A. Meissner$^{1,2}$ and Hermann Nicolai$^3$}
\affiliation{$^1$ Faculty of Physics,
University of Warsaw\\
Ho\.za 69, Warsaw, Poland\\
$^2$ National Centre for Nuclear Research, \'Swierk, Poland\\
$^3$ Max-Planck-Institut f\"ur Gravitationsphysik
(Albert-Einstein-Institut)\\
M\"uhlenberg 1, D-14476 Potsdam, Germany\\
}

\begin{abstract} We show how neutrino mixing leads to a softening of the UV behavior
of neutrino loops, and can give rise to anomaly-like, but manifestly UV finite amplitudes.
This mechanism may be of special significance for the coupling of dark matter-like
particles to the Standard Model, as we illustrate with a one-loop example involving
the coupling to $W$-bosons.

\end{abstract}
\pacs{14.60.Pq,14.60.St,14.80.Va}

\maketitle

\vspace{0.2cm}

In this letter we explain a mechanism specific to neutrino loop diagrams in the
minimally extended Standard Model (SM) with right-chiral neutrinos. This effect
not only improves the UV behavior of loop amplitudes but, more importantly,
can lead to anomaly-like amplitudes and effective couplings in processes involving
dark matter-like particles.  Although elementary  (and used implicitly in our previous work \cite{MN}) this mechanism has not been spelled out explicitly so far in the existing literature to the best of our knowledge. It relies crucially on neutrino mixing, and on the simultaneous presence of Majorana- and Dirac-like mass and Yukawa terms for the right-chiral fields. While reminiscent of  Pauli-Villars regularization, it involves only real particles, and no ghosts  appear. Remarkably, this appears to be the only context where anomaly-like effects can arise from triangle diagrams {\em without} the linear UV divergences that  one commonly associates with anomalous amplitudes.

The relevant diagrams typically involve amplitudes with `dark matter' (axion-like) particles on one side of the diagram, and `visible' SM particles on the other. Our results pertain in particular to scenarios  where the `heavy' neutrinos are not so heavy after all \cite{MN,MS}, in contradistinction to the more common see-saw scenarios with Majorana mass scales of $\cO( 10^{10} \,{\rm GeV})$ or more \cite{seesaw}. For these we arrive at amplitudes which are not only UV and IR finite, but also can explain in a perfectly natural manner  the extreme smallness of dark matter couplings to SM particles, and this without any fine-tuning or the need to adduce  the kind of `shadow matter' invoked by numerous other currently popular  scenarios.

For simplicity, we consider a one family model  with right-chiral neutrinos.  Using Weyl spinors to express the 4-component neutrino spinor
$\cN \equiv (\cN_L, \cN_R)\equiv (\nu_\al , \bar{N}^{\dot\al} )$ and its conjugate in terms of $SL(2,\mathbb{C})$ spinors, the neutrino part of the kinetic Lagrangian reads
\bea
 \cL &=& \frac{\ri}{2}\left(\nu^{\al} \dslash_{\al\dot\be}\bar\nu^{\dot\be} +
 %\bar\nu_{\dot\al}\dslashbar^{\dot\al\be}\nu_\be +
 N^{\al} \dslash_{\al\dot\be}\bar N^{\dot\be}
 %\bar{N}_{\dot\al}\dslashbar^{\dot\al\be}N_\be\right) \nn \\
+ {\rm c.c.}\right) - \\
&& - m\nu^{\al}N_\al -m^*\bar\nu_{\dot\al}\bar{N}^{\dot\al} - \frac{M}{2} N^{\al}N_\al -\frac{M^*}{2}\bar{N}_{\dot\al}\bar{N}^{\dot\al}  \nn
\label{kinterms}
\eea
where we have included both Dirac and Majorana mass terms (for the two-component
formalism see e.g. \cite{BW}). In the SM these are supposed to arise via spontaneous symmetry breaking, such that $m= Y_\nu \langle H\rangle$ and $M= Y_M \langle\phi\rangle$ with appropriate Yukawa couplings $Y_\nu$ and $Y_M$, and non-vanishing vacuum expectation values for the SM Higgs field $H(x)$ and a  further complex (electroweak singlet) scalar field $\phi (x)$~
 \footnote{Recall that the neutrino mass hierarchy can be implemented by taking $Y_\nu \sim \cO(10^{-5})$ \cite{MN,MS}, that is, {\em without} introducing a
  large intermediate mass scale by hand. In this case, the heavy neutrinos typically assume
  mass values in a range {\em below $\cO(1\,{\rm TeV})$}. }.
We assume $m$ and $M$ real,  as this can be  achieved by appropriate phase rotations.

The standard procedure to deal with (\ref{kinterms}) consists in diagonalizing the mass matrix, introducing new fields
\beq
\begin{bmatrix} \nu' \\ N' \end{bmatrix} = \begin{bmatrix} \cos \th & -\sin \th \\ \sin \th  & \cos\th \end{bmatrix} \begin{bmatrix} \nu \\ N \end{bmatrix} \label{combination}
\eeq
in terms of which the kinetic terms become diagonal
\bea
 \cL &=& \frac{\ri}{2}\left(\nu'^{\al} \dslash_{\al\dot\be}\bar\nu'^{\dot\be} + N'^{\al} \dslash_{\al\dot\be}\bar N'^{\dot\be} + {\rm c.c}\right) -   \\
&& -\frac{m'}{2}\left( \nu'^{\al}\nu'_\al + \bar\nu'_{\dot\al}\bar\nu'^{\dot\al}\right) - \frac{M'}{2} \left( N'^{\al}N'_\al + \bar{N}'_{\dot\al}\bar{N}'^{\dot\al}\right)\nn
\eea
A simple calculation gives
\beq
\tan 2\th = \frac{2m}{M}
\eeq
and the mass eigenvalues $m'$ and $M'$
\beq
m' =  -\frac{\sin^2\th}{\cos 2\th}\,M,\ \ \ \
M' = \frac{\cos^2\th}{\cos 2\th}\,M
\label{mM'}
\eeq
Notice that $m'M' < 0$, so $m'$ and $M'$ always come with {\em opposite signs}.
The propagators then take the standard diagonal form,
\bea
\langle \nu'_\al(x) \bar \nu'_{\dot\be}(y) \rangle &=& \ri\int\frac{d^4p}{(2\pi)^4}\frac{\pslash_{\al\dot\be} }{p^2-m'^2} \re^{-\ri p\, (x-y)} \label{prop1}\\
\langle \nu'_\al(x) \nu'^\be(y) \rangle &=& \ri\int\frac{d^4p}{(2\pi)^4}\frac{m' \de_\al^{\;\be}}{p^2-m'^2} \re^{-\ri p\, (x-y)} \label{prop2}
\eea
with analogous expressions for the $N'$ propagators after replacing $m'\to M'$. With the redefinitions (\ref{combination}) the SM interaction vertices involving neutrinos are now {\em off-diagonal}, and any Feynman diagram computation will involve a sum of contributions for each vertex. We will refer to this description (diagonal propagators, off-diagonal vertices) as the `propagation picture', because it is more natural if we ask about the eigenstates of propagation.

As shown in our previous work \cite{LMN}, however, there is another description (which we refer to as the `vertex picture') where the vertices remain diagonal, while instead the propagators are off-diagonal. This picture is better adapted to the fact that the weak interactions in the SM involve only left-chiral particles. In that picture all diagrams with both heavy and light neutrinos circulating in the loops are manifestly {\em UV finite}, due to the unusual fall-off
properties of the propagator components for large momenta. More specifically, the
propagators are determined by inverting the kinetic matrix
%\beq
%\cL = \frac12 \begin{bmatrix}\nu_\al & \bar\nu_{\dot\al} & N_\al & \bar N_{\dot\al}\end{bmatrix}
%\begin{bmatrix}\nu_\be \\ \bar\nu_{\dot\be} \\ N_\be \\ \bar N_{\dot\be}\end{bmatrix}
%\eeq
%where the matrix $A$ is equal to
\beq
A=
\begin{bmatrix}
0 & \ri\dslash^{\al\dot\be} & m\eps^{\al\be} & 0 \\
\ri\dslashbar^{\dot\al\be} & 0 & 0 & -m\eps^{\dot\al\dot\be} \\
m\eps^{\al\be} & 0 & M\eps^{\al\be} & \ri\dslash^{\al\dot\be} \\
0 & -m\eps^{\dot\al\dot\be} & \ri\dslashbar^{\dot\al\be} & -M\eps^{\dot\al\dot\be}
\end{bmatrix}
\eeq
with the result (in momentum space)
\bea
\langle \nu_\al \nu_\be \rangle &=& \ri Mm^2 \cD(p) \eps_{\al\be}  \nn\\
\langle \nu_\al \bar\nu_{\dot\be} \rangle &=& \ri \left(p^2- M^2 - m^2\right)\cD(p) \pslash_{\al\dot\be} \nn\\
\langle N_\al N_\be \rangle &=& \ri M p^2\cD(p) \eps_{\al\be}  \\
\langle N_\al \bar{N}_{\dot\be} \rangle &=& \ri\left(p^2-m^2\right)\cD(p) \pslash_{\al\dot\be} \nn\\
\langle \nu_\al N_\be \rangle &=& \ri m\left(p^2- m^2 \right)\cD(p) \eps_{\al\be}\nn\\
\langle \nu_\al \bar{N}_{\dot\be} \rangle &=& -\ri mM \cD(p) \pslash_{\al\dot\be}\nn
\eea
where $ \cD(p):= \big[ (p^2)^2 - p^2(2m^2 + M^2) + m^4 \big]^{-1}$.
In this description the finiteness of the neutrino loop diagrams is a
consequence of the fall-off properties of the propagators
$\langle \nu_\al \nu_\be \rangle$  and $\langle \nu_\al \bar{N}_{\dot\be}\rangle$
which both decay faster than $1/p^2$.

Alternatively, this effect can be understood in the `propagation picture',
where the propagators have the usual fall-off properties. Namely, the
cancellations are now due to the fact that off-diagonal vertices give rise to
several propagator contributions that must be summed over for a given diagram,
and thus follow from the formula relating the $\nu$, $N$ and
$\nu'$, $N'$ propagators via (\ref{combination}),
\bea
\langle \nu_\al \bar N_{\dot\be} \rangle &=& \sin\th\cos\th \left(- \langle \nu'_\al \bar \nu'_{\dot\be} \rangle +  \langle N'_\al \bar N'_{\dot\be} \rangle \right) \label{1stcancel} \\
\langle \nu_\al \nu^\be \rangle &=& \cos^2\th \langle \nu'_\al \nu'^\be \rangle + \sin^2\th \langle N'_\al N'^\be \rangle
\label{2ndcancel}
\eea
Cancellation of the leading terms for large $|p|$ in (\ref{1stcancel}) is obvious, while cancellation in (\ref{2ndcancel}) is due to the relation $m'\cos^2 \theta + M'\sin^2\theta = 0$, that is, happens thanks to the opposite signs of $m'$ and $M'$, cf. (\ref{mM'}). In both cases the cancellation is thus due to the fact that there are two propagator components contributing {\em with opposite signs} in the loop,  just like in Pauli-Villars regularization.

Let us illustrate this mechanism with a simple example. The SM with right-chiral fermions and spontaneously broken $SU(2)_w\times U(1)_Y$ symmetry in particular contains the following vertices (cf. for example \cite{Pok})
\bea
\cL_{\text{int}} &\supset& - \frac{g_W}{\sqrt{2}}W_\mu^+ e_L^{\al}\si^\mu_{\al\dot\be}\bar{\nu}^{\dot\be} - \frac{g_W}{\sqrt{2}}W_\mu^- \nu^{\al}\si^\mu_{\al\dot\be}\bar{e}_L^{\dot\be} + \nn \\
&& - \frac{\ri M}{2\sqrt{2}\langle\varphi\rangle} \, a (N^{\al}N_\al -  \bar N_{\dot\al} \bar N^{\dot\al})
\eea
where the Yukawa interaction arises from a Majoron-like coupling
$\frac12 Y_M (\phi N^\al N_\al + {\rm c.c.})$  \cite{CMR} upon expanding
\beq
\phi(x) = \varphi(x) \exp\big(\ri a(x)/\sqrt{2}\langle\varphi\rangle\big) = \langle\varphi\rangle
+ \frac{\ri}{\sqrt{2}} a(x)  + \cdots \;.
\eeq
Consequently, the field $a(x)$ carries the charge of a spontaneously broken global
symmetry, and therefore couples like a Goldstone boson (in \cite{MN} this symmetry
is lepton number symmetry). In this model, we now consider the one-loop
$aW^+ W^-$ amplitude, shown in Fig. 1a, which is a sum of two diagrams with
either  $\nu N$ or $\nu \bar{N}$ propagators in the loop in the `vertex picture'. It reads
\bea
&& -\ri\cM^{\mu\nu}_{aWW}(p,q) = \nn \\
&=& \frac{g_W^2 M}{2\sqrt{2}\langle\varphi\rangle} \int\frac{d^4k}{(2\pi)^4} \left\{ \si^\mu_{\al_1\dot\be_1} \langle \bar{e}_L^{\dot\be_1}e_L^{\al_2} (k-p) \rangle \si^\nu_{\al_2\dot\be_2}  \times \right. \nn \\
&& \times \left[ \langle \bar{\nu}^{\dot\be_2}N^{\al_3} (k) \rangle  \langle N_{\al_3}\nu^{\al_1} (k+q) \rangle \right. + \nn \\
&& \qquad - \left.\left.\langle \bar{\nu}^{\dot\be_2}\bar{N}_{\dot\al_3} (k) \rangle \langle \bar{N}^{\dot\al_3}\nu^{\al_1} (k+q) \rangle \right]\right\} \\
&=& \frac{\ri g_W^2}{2\sqrt{2}\langle\varphi\rangle }\int\frac{d^4k}{(2\pi)^4}  \frac{(k-p)_\la}{(k-p)^2-m_e^2}\,  \text{Tr}\{\si^\si\bar{\si}^\mu\si^\la\bar{\si}^\nu \}
\times\nn\\
&& \!\!\!\!\!\!\frac{m^2 M^2 [-k_\si (2kq+q^2)+ q_\si (k^2-m^2)]}{\{(k^2-m^2)^2-k^2 M^2\}\{[(k+q)^2-m^2]^2-(k+q)^2 M^2\}} \nn
\eea
For small axion momentum $q^\mu$,  it suffices to retain only the leading terms
in $q^\mu$ and $m'$, so we get
\bea
&& -\ri\cM^{\mu\nu}_{aWW}(p,q)
\approx -\ri q^\rho \frac{g_W^2 m' M'^3}{2\sqrt{2}\langle\varphi\rangle }
\,  \text{Tr}\{\si^\si\bar{\si}^\mu\si^\la\bar{\si}^\nu \}
\times \nn\\
&& \ \ \times\int\frac{d^4k}{(2\pi)^4}  \frac{(k-p)_\la}{(k-p)^2-m_e^2} \frac{(2k_\si k_\rho- g_{\si\rho} k^2)}{[k^2(k^2-M'^2)]^2}
\eea
where we have now substituted the mass eigenvalues $m'$ and $M'$ for $m$ and $M$.
Since the integrand decays as $\sim |k|^{-7}$ for large $k^2$ the integral is UV convergent. There is no IR divergence either, which justifies taking the limit $m'\rarr 0$ in the denominator.

The similarities with the standard anomalous triangle diagram are evident, but we
here arrive at a perfectly finite expression, without the need of any regularization. Because
\bea
\text{Tr}\{\si^\si\bar{\si}^\mu\si^\la\bar{\si}^\nu \} &=& \nonumber\\
&& \!\!\!\!\!\!\!\!\!\!\!\!\!\!\!\!\!\!\!\!\!\!\!\!\!\!\!\!\!\!\!\!\!\!\!\!
= 2\left(\eta^{\si\mu} \eta^{\la\nu}-\eta^{\si\la}\eta^{\mu\nu}+\eta^{\si\nu}\eta^{\mu\la}-
\ri\eps^{\si\mu\la\nu}\right)\nonumber
\eea
the result contains both `parity-even' and `parity-odd' contributions, reflecting the fact that parity is maximally broken in the SM.  In particular, we do get a contribution $\propto\eps^{\sigma\mu\lambda\nu}$ mimicking the anomaly.

Using Feynman parameters, we can perform the integral over momenta and get
\bea
&& -\ri\cM^{\mu\nu}_{aWW}(p,q) \approx
q^\rho \frac{g_W^2 m' M'^3}{32\pi^2\sqrt{2}\langle\varphi\rangle } \,
%a^{\si\mu\la\nu}
\times\label{res} \nn\\
&&  \text{Tr}\{\si^\si\bar{\si}^\mu\si^\la\bar{\si}^\nu \}  \times
\int_0^1 dy\int_0^1 dx\, x(1-x)y^3 \times \nonumber \\
&& \times \left[ \frac{[(-1+2y) p_\la g_{\rho\si} +(1-y)(p_\rho g_{\si\la}+p_\si g_{\rho\la})]}{[-y(1-y)p^2+(1-y)m_e^2+yxM'^2]^2} +\right. \nn \\
&& \qquad\qquad \left. +\frac{2y(1-y)^2 p_\la (-p^2g_{\rho\si} + 2p_\rho p_\si)}{[-y(1-y)p^2+(1-y)m_e^2+yxM'^2]^3}\right]  \nn
\eea
For small $p^2$ we obtain ($\al_W \equiv g_W^2/{4\pi}$, $\eta_e \equiv m_e/M$)
\bea
&& -\ri\cM^{\mu\nu}_{aWW}(p,q) \approx \\
&\approx& -q^\rho \frac{\al_W m'}{48\pi\sqrt{2}\langle\varphi\rangle M' }
\,  \text{Tr}\{\si^\si\bar{\si}^\mu\si^\la\bar{\si}^\nu \}
\times \nn \\
&& \times\left[p_\la g_{\rho\si} \frac{(1-5\eta_e^2)
\ln \eta_e^2 -(1+3\eta_e^2)(1-\eta_e^2)}{(1-\eta_e^2)^3}\right. + \nonumber\\
&& \quad \left. + (p_\rho g_{\si\la}+p_\si g_{\rho\la}) \frac{(1+\eta_e^2)\ln \eta_e^2 + 2(1-\eta_e^2)}{(1-\eta_e^2)^3} \right]\nn
\eea
(the singularity for $\eta_e=1$ is spurious). For large $p^2$
\bea
&& -\ri\cM^{\mu\nu}_{aWW}(p,q) \approx \frac{\al_W m' M'}{8\pi\sqrt{2}\langle\varphi\rangle }\,
\left[1+O\left(\frac{\ln (p^2/M^2}{p^2}\right)\right]\nn\\
&& \quad \times \, \frac{2}{p^2}\, \big[ p^\mu q^\nu + p^\nu q^\mu - \eta^{\mu\nu} p\cdot q -
\ri \varepsilon^{\mu\nu\rho\lambda} p_\rho q_\lambda \big]
\eea
where we have written out the trace. The result contains the manifestly non-gauge
invariant pieces corresponding to the first three terms in brackets, which are expected in accordance with the fact that the electroweak $SU(2)_w\times U(1)_Y$ symmetry is broken
 ~\footnote{This link is somewhat indirect,  but readers should keep in mind
  that without spontaneous symmetry breaking this
  diagram would be absent altogether.}. 
In addition, there is an anomaly-like contribution
$\propto a \, \eps^{\mu\nu\rho\sigma} \partial_\mu W^+_\nu \partial_\rho W^-_\sigma$.
The full amplitude falls off linearly in $p^\mu$, whence this `form factor' will also improve the UV behavior of higher loop diagrams when inserted as a
subdiagram \cite{LMN}.

Because $a(x)$ is a Goldstone boson, one momentum factor can always be pulled out
of the diagram. With unbroken gauge invariance, this would only leave the `anomaly-like', but in this case manifestly UV finite amplitude, whereas with external $W$ or $Z$ bosons there also appear the manifestly non-gauge invariant contributions displayed above. By contrast, for higher loop diagrams with external gluons of the type considered in \cite{LMN}, only the anomaly-like interaction can survive, because $SU(3)_c$ remains unbroken, and the Chern-Simons-like coupling
$\partial_\mu a \, \cJ^\mu_{CS} \cong  a \, {\rm Tr}\, \tilde{G}G$ is the only possibility to reconcile gauge invariance with the Goldstone property of $a(x)$. Similar comments apply to the coupling of $a(x)$ to photons, which comes out proportional to $a\tilde{F}F$.
This explains why in both cases the amplitude acquires an `axion-like'  form.

We conclude with three remarks. First of all, similar results and conclusions are
obtained for the analogous diagram with external $Z$-bosons shown in Fig.~1b,
where the triangle loop is `purely neutrino'. Secondly, the mechanism exhibited here persists in more realistic scenarios with three families, as is most easily seen in
the `vertex picture'. In this case, we have correspondingly more complicated  mass matrices,
\beq
\cM=
\begin{bmatrix}
0 & 0 & m\, \eps^{\al\be} & 0 \\
0 & 0 & 0 & -m^*\eps^{\dot\al\dot\be} \\
m^T\eps^{\al\be} & 0 & M\eps^{\al\be} & 0 \\
0 & -m^\dagger \eps^{\dot\al\dot\be} & 0 & -M\eps^{\dot\al\dot\be}
\end{bmatrix}
\eeq
where $m$ is now a {\em complex} 3-by-3 matrix, and $M$ a real diagonal 3-by-3 matrix. With the usual see-saw assumption of a large hierarchy between $m$ and $M$ the heavy neutrino masses are given approximately by the eigenvalues of $M$, while the squared mass  eigenvalues of the light neutrinos follow by diagonalizing the hermitean matrix $M^{-1/2}m^\dagger m M^{-1/2}$ (for a recent discussion of corrections to these formulae see e.g. \cite{ML}).

Finally, in \cite{MN} we have proposed to actually {\em identify} the field $a(x)$
(usually referred to as the `Majoron') with the axion, in which case the smallness
of various axion couplings can be explained very naturally, and without the need to
invoke any large intermediate mass scales. This proposal is based on the
fact that the effective couplings of $a(x)$ to gluons and photons assume the standard axionic form. In this Letter, we have provided a direct argument explaining why this happens. We hope that the new effect exhibited here will lend more credence to this proposal, which in our opinion possesses attractive features in comparison to other axion models.

\vspace{0.5cm}

\onecolumngrid
\begin{figure}[ht]
\hspace{-1cm}
\begin{minipage}[b]{0.3\linewidth}
\centering
\includegraphics[scale=0.8,viewport=0 630 840 750,clip]{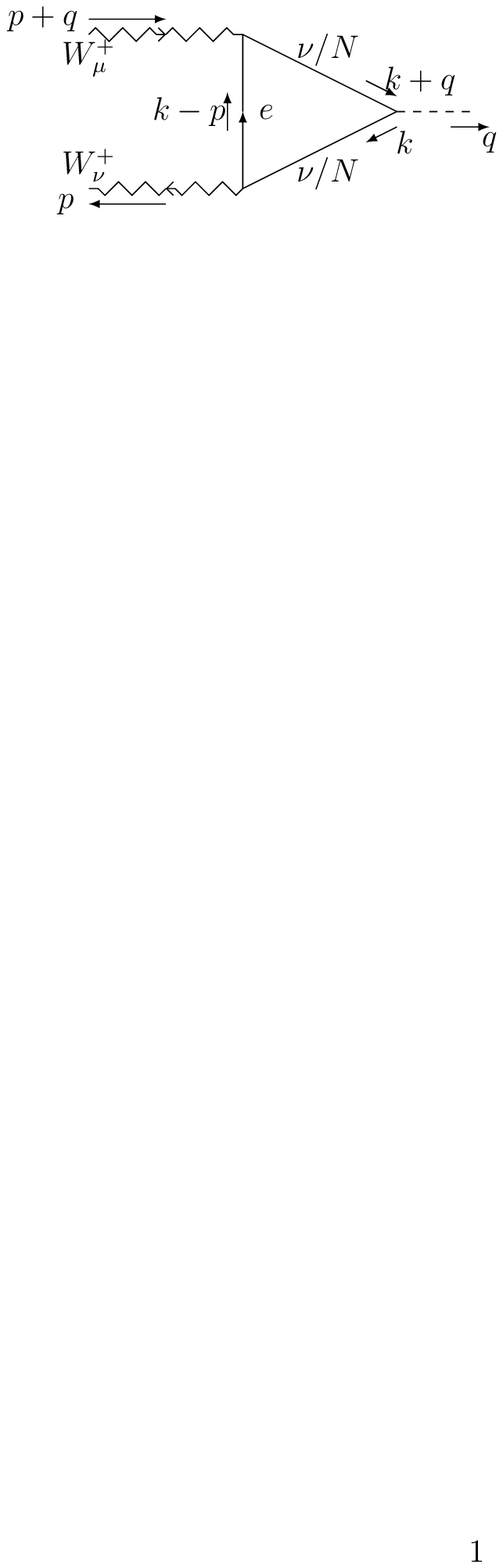}
\label{fig:figure1}
\end{minipage}
\hspace{3.5cm}
\begin{minipage}[b]{0.3\linewidth}
\centering
\includegraphics[scale=0.8,viewport=0 630 1840 750,clip]{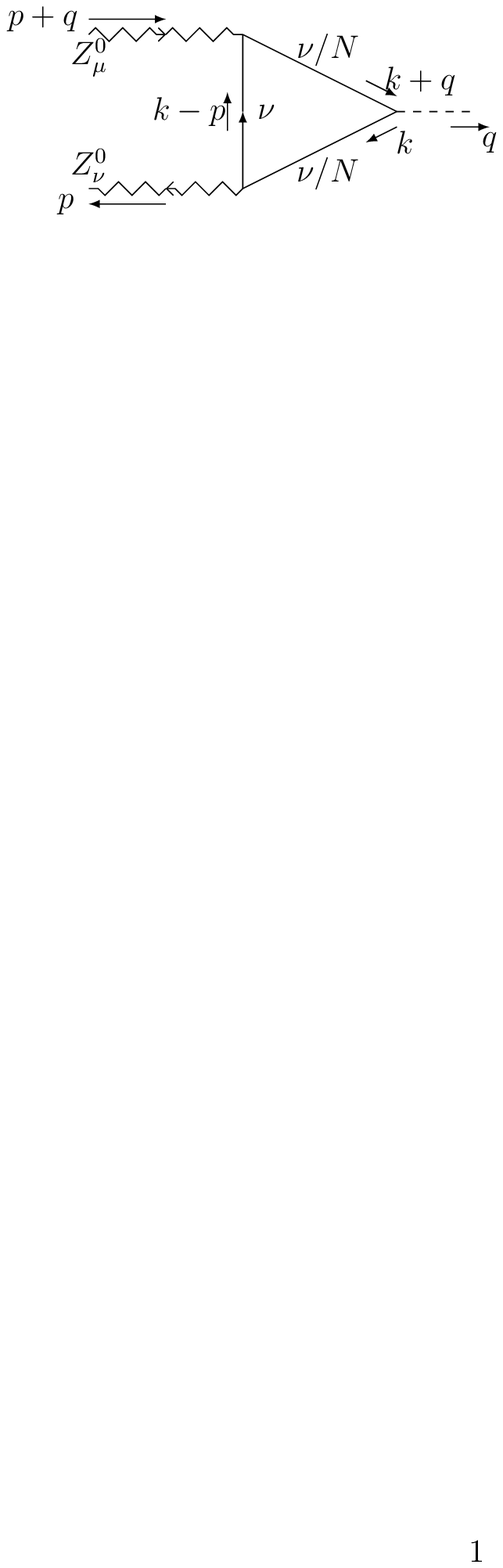}
\label{fig:figure2}
\end{minipage}
\end{figure}
\begin{center}
Fig 1. a) a-W-W diagram b) a-Z-Z diagram
\end{center}

\end{document}